# From the Lab to the Street:
## *Solving the Challenge of Accelerating Automated Vehicle Testing*

**DING ZHAO, PhD**
Assistant Research Scientist
Mechanical Engineering
University of Michigan

**HUEI PENG, PhD**
Director, Mcity
Roger L. McCarthy Professor
of Mechanical Engineering
University of Michigan

**Contents**

**EXECUTIVE SUMMARY**

As automated vehicles and their technology become more advanced and technically sophisticated, evaluation procedures that can measure the safety and reliability of these new driverless cars must develop far beyond existing safety tests. To get an accurate assessment in field tests, such cars would have to be driven millions or even billions of miles to arrive at an acceptable level of certainty – a time-consuming process that would cost tens of millions of dollars.

Instead, researchers affiliated with the University of Michigan's Mcity connected and automated vehicle center have developed an accelerated evaluation process that eliminates the many miles of uneventful driving activity to filter out only the potentially dangerous driving situations where an automated vehicle needs to respond, creating a faster, less expensive testing program. This approach can reduce the amount of testing needed by a factor of 300 to 100,000 so that an automated vehicle driven for 1,000 test miles can yield the equivalent of 300,000 to 100 million miles of real-world driving.

While more research and development needs to be done to perfect this technique, the accelerated evaluation procedure offers a ground-breaking solution for safe and efficient testing that is crucial to deploying automated vehicles.

**THE PROBLEM**

The dawning of driverless vehicles presents a number of challenges for automakers, regulators and city planners, from the design of software and hardware in the vehicles, to redesigning the road infrastructure, to clarifying the challenging legal issues about potential liability in an accident.

But before consumers will embrace automated vehicles – especially cars with no driver controls at all – the people who will buy and ride in these "cars of the future" will need to be assured that the vehicles are reliable and safe.

Safety testing in today's cars and trucks is a well-defined, standardized effort: For crashworthiness, get your test vehicle, install the crash-test dummies and sensors, put it on a test sled, roll the video cameras and see what happens when the car hits the wall. For rollover vulnerability, conduct a few well defined steering maneuvers, and compute a rollover "score" using results of the vehicle-in-motion test and taking into account the vehicle's shape and weight distribution. The results are easily measured and can be repeated in a way that assures car buyers, government regulators, and insurance companies.

The crashworthiness test measures the outcome of a single event: What happens when a car crashes at a particular speed in a particular way and how badly are the occupants hurt? The rollover tests rate the propensity for a tip-over. But gauging with any kind of certainty how an automated vehicle will react is vastly more difficult than looking to see whether the crash-test dummy's arm got broken. Test methods for traditionally driven cars are something like having a doctor take a patient's blood pressure or heart rate, while testing for automated vehicles is more like giving someone an IQ test. The variables of traffic and road conditions, weather, time of day, and the unpredictable actions of other drivers and vehicles present a constantly changing tangle of variables that an automated car will need to recognize and process to make the right, safe choice.

In fact, even the question of how to design such tests is much more complicated. Instead of, "What happens in a crash?" the tests for automated vehicles must measure how effectively these cars can keep one from happening.

Adding to the challenge is the fact that, as automated vehicles are introduced, they won't start off dominating the road. Instead, the new driverless cars will share the road for years to come with vehicles driven by humans. While connected, automated cars will be able to talk to each other to avoid crashes, a driverless car won't hear a peep out of that 2006 pickup that's about to veer into the next lane because the driver just spilled coffee in his lap.

All these factors pose a huge challenge to manufacturers: How to develop tests for automated vehicles that can accurately represent and replicate unpredictable, wildly varying real-world driving situations. It's a problem that far exceeds anything vehicle test engineers have faced before. The old test matrix based assessment process, with pre-defined testing scenarios, simply no longer applies. Something radically different and innovative is needed.

**APPROACH**

To create consumer acceptance of automated vehicles, tests will need to prove at a level of 80 percent confidence that the robotic vehicle is 90 percent safer than human drivers on the road. The distance test vehicles would need to be driven in simulated or real-world settings to get to that high confidence level would be 11 billion miles.

And, to be truly safe, a robotic vehicle would need to be able to properly respond to the most dangerous driving situations, which turn out to be pretty rare. According to the National Highway Traffic Safety Administration, an accident serious enough to be reported to police – typically, one with at least $1,000 worth of vehicle damage – occurs once in just every 530,000 miles of driving. A crash that results in a fatality is even rarer – once in every 100 million miles of driving.

Now consider that the typical driver clocks about 12,000 miles per year. In an urban environment where driving conditions are more complex, speeds range from 10 mph to 25 mph in congested traffic, meaning that a test driver putting in an 8-hour shift wouldn't be able to collect data covering more than 200 miles, at best. At that rate, it would take more than 27 years just to get to 2 million miles – an impressive feat but still extremely short of what's needed. Up the testing to three shifts covering 24 hours a day, and you'd still need about 3,300 days of driving – more than 9 years to reach 100 million miles. That's a lot of drivers, a lot of gas, and a lot of vehicles and repairs. And even then, the amount of data on significant events will be slim because, based on crash statistics, we know that researchers get an interesting and useful piece of incident data roughly once every 100,000 miles of driving.

That means even the most advanced and large-scale efforts to test automated vehicles today fall woefully short of what is needed to thoroughly test these robotic cars.

To address this problem, U-M engineers set out to adapt the concept of "accelerated longitudinal evaluation," which is already widely used in the auto industry. Consider corrosion testing: Car makers don't set cars outside for 10 years to see if the rocker panels rust out. Instead, they use concentrated solutions of chloride and varying conditions of relative humidity on a test track of salt troughs, mud troughs and gravel roads to speed-up any potential rust. Once the engineers obtain the equivalent to a year's worth of exposure, a standard calculation of corrosion rates predicts rust resistance over time.

A similar technique can be applied to testing automated vehicles, according to groundbreaking research conducted by Ding Zhao, PhD, assistant research scientist in Mechanical Engineering at U-M, working with Huei Peng, PhD, the Roger L. McCarthy Professor of Mechanical Engineering at U-M, and director of Mcity. The key is to break down difficult real-world driving situations into components that can be tested or simulated repeatedly. Two scenarios have been tested: car-following and merging/cut-in. In both cases, the tested automated vehicle is the car behind. It responds to the lead vehicle maneuver, which simulates the behavior of a human-controlled vehicle.

Adding to the testing challenge is the fact that, for higher-level automated vehicles, the evaluating maneuvers need to be much more sophisticated. In Level 1 and Level 2 automated vehicles, human drivers handle the actual monitoring and driving, with automated systems assisting steering, acceleration and braking. Such systems exist now, as do methods for evaluating them.

But as the level of automation increases, the robot driver needs to handle a much wider set of scenarios. At Level 4 and Level 5 automation – the two most advanced – automated systems control the driving and response of the car. A Level 4 vehicle, for example, must be able to contend with navigating left turns in front of traffic, avoiding bicyclists and pedestrians and merging onto highways, as well as all lower-level driving events, even if a human driver is in the car but doesn't respond when prompted.

Then consider that each manufacturer has its own distinct approach to designing, building, and programming automated vehicle control systems. To protect their proprietary systems, manufacturers can't disclose the technical details, leaving researchers able to test only the outcomes from what is essentially a secretive black box controlling each different type of vehicle.

This presents a considerable challenge to the four basic approaches to vehicle testing:

**Naturalistic Field Operational Tests:** This is driving in real-world or simulated real-world conditions, and produces data on driver performance, behavior, environment, driving context and other factors that were associated with critical incidents, near misses and crashes. The drawback to this approach is that it requires a lot of vehicles, and is time-consuming and expensive. The average driver would need to be behind the wheel for 38 years to produce one significant crash that can provide data, while earlier research pegged the cost of field testing projects to be at least $10 million to generate statistically meaningful results.

**Test Matrix:** This approach presents a number of pre-defined scenarios that each vehicle goes through for evaluation. For example, a test of automated emergency braking uses three different scenarios, including one where the car faces a stopped vehicle, another where the car in front keeps a steady speed, and a third where the car in front brakes to slow down. The matrix approach can be used in field tests and in simulations. Problems with this approach stem from the fact that the scenarios all are predefined and predetermined, and the tests are largely designed based on data from human drivers rather than automated vehicles.

**Worst-Case Scenario:** As the name implies, the most serious driving scenarios and parameters are selected. This is a good approach for identifying weaknesses in the design of the vehicle being tested. However, it doesn't accurately assess risk or probability in real-world situations. Also, the worst case for one vehicle system might not be the worst case for another. Similarly, different automated vehicles will be challenged by different worst-case maneuvers. In other words, this test procedure cannot be used for government standardized testing, or to infer the expected safety performance of an automated vehicle, and it cannot be used to determine a fair insurance rate. It is, however, a useful tool for a company to understand the worst-case vulnerability of its automated vehicle, possibly leading to design changes.

**Monte Carlo Simulation:** First developed for the Manhattan Project, this approach allows for a mathematical assessment of risk and probability in a wide range of outcomes, calculating over and over again using different probabilities and potential outcomes. Using scenarios from real-world driving data, however, means that uneventful driving outcomes will be evaluated more often, reducing the efficiency of the testing.

While each of these four evaluation approaches presents certain benefits, each one also comes with drawbacks that mean the results either won't represent real-world driving conditions, or that the results won't accelerate the pace of testing. What's needed instead is an accelerated evaluation process that can distill potentially dangerous vehicle interactions into a compressed test that still accurately reflects what actually happens on the road statistically. By stripping out the long stretches of uneventful driving, when an automated vehicle won't need to react to a threat, the evaluation process can be made faster and cheaper.

To develop the accelerated evaluation process, researchers began with a six-step analysis of driving data:

- Collect a large amount of data from real-world driving;
- Distill this data down to only those events that can contain meaningful interactions between an automated vehicle and a vehicle piloted by a human driver;
- Model the behaviors of vehicles piloted by a human driver as the major threat to automated vehicles as random variables with a distribution of probability;
- Reduce the non-safety-critical parts of daily driving and replace them with increased occurrences of critical events;
- Run Monte Carlo tests with the accelerated scenarios to create more intense interactions/crashes between automated and human-driven vehicles;
- Use statistical analysis to mathematically reverse the accelerated test results to see how the automated vehicle would perform in everyday driving conditions statistically.

The driving data for this analysis was collected by the University of Michigan Transportation Research Institute in the Safety Pilot Model Deployment Program and the Integrated Vehicle-Based Safety Systems Program, conducted in southeast Michigan and near the main U-M campus in Ann Arbor, Michigan. The Integrated Vehicle program included 16 light vehicles operated by 108 volunteer drivers for 6 weeks, which collected a data set representing 213,309 miles, 22,657, and 6,164 hours of driving. The Safety Pilot program involved more than 2,800 vehicles and data was collected from August 2012 to June 2014, covering almost 4 million trips that traveled more than 25 million miles in nearly 900,000 hours. The Safety Pilot vehicles were equipped to transmit and to receive driving data from other connected vehicles and connected infrastructure elements.

After the six-step analysis of the problem, researchers developed four methodologies that form the basis of the U-M accelerated evaluation process to rapidly speed the testing of automated vehicles.

The first method is based on how frequently a significant driving event happens on the road, and strips out the more common, uneventful safe driving situations. The second uses importance sampling to statistically increase the number of critical driving events in a way that still accurately reflects real-world driving situations. The third method is to construct a formula that accurately distills those critical events, test it, and then apply it to further reduce the amount of testing required. Finally, the interactions between human-driven vehicles and robotic vehicles is analyzed based on optimizing the random occurrences of significant driving events in the most complex scenarios.

The accelerated analysis research was conducted on the two most common situations resulting in serious crashes. The first was where the automated vehicle was following one driven by a human, where adjustments constantly must be made for movements of the lead vehicle, as well as speed, road and weather conditions and other rapidly changing factors. The second involved a human-driven car cutting in front of the automated car that was being followed, in turn, by another human-driven vehicle. Three metrics – crash, injury, and conflict rates – were calculated, along with the likelihood that one or more passengers in the automated vehicle would suffer moderate to fatal injuries. The accuracy of the evaluation was determined by conducting and then comparing accelerated and real-world simulations.

**CONCLUSIONS**

By combining all four methodologies in one over-arching process, the result is U-M's accelerated evaluation procedure, which can cut the time required to evaluate crash, injury, or other conflict events by 300 to 100,000 times. If an automated vehicle drives for 1,000 miles under this method that exposes it to a condensed set of the most serious and challenging driving situations, it would yield the equivalent of 300,000 to 100 million miles of real-world driving.

This accelerated testing approach can eliminate up to 99.9 percent of the cost and time in compiling enough data to achieve a level of 80 percent confidence that any such tested robotic vehicle is 90 percent safer than cars piloted by human drivers now on the road. This new evaluation process would dramatically reduce the amount of time and cost involved in validating the reliability of automated vehicles.

To achieve that level of confidence, evaluators will need many more miles of real-world driving with an automated vehicle. That means the current amount of data about the real-world driving situations to which Level 4 robotic vehicles will need to accurately respond isn't nearly enough.

In addition, researchers also will need to identify more critical driving scenarios to analyze all the potential failures of automated vehicles, including challenges to sensors from snow and fog; blinking signal lights or gestures from other drivers; illegal movements, such as vehicles running red lights or jaywalking pedestrians; movements by heavily loaded vehicles that behave and respond differently; and various road conditions.

Finally, U-M researchers also aim to expand evaluations to three more critical driving situations beyond car-following and lane changes, to include left turns, street crossing and cars coming in the opposite direction. They also want to include scenarios for single-vehicle crashes and accidents involving pedestrians and cyclists.

Once that data and expanded evaluation capability is developed, U-M researchers will refine their discoveries in accelerated evaluation, so that this innovative methodology can be employed across a wide variety of vehicles and technology to show consumers that automated vehicles are safe and trustworthy.

## RESOURCES

*About Mcity*
*Mcity at the University of Michigan is leading the transition to connected and automated vehicles. Home to world-renowned researchers, a one-of-a-kind test facility, and on-road deployments, Mcity brings together industry, government, and academia to improve transportation safety, sustainability, and accessibility for the benefit of society.*